\documentclass[12pt]{article}

\catcode`\@=11

\global\arraycolsep=2pt
\usepackage{amsbsy,amssymb,latexsym,amsfonts,amsmath}
\usepackage{graphicx,color}

\allowdisplaybreaks

\begin{document}
\begin{flushright}
\parbox{4.2cm}
{RUP-21-15}
\end{flushright}

\vspace*{0.7cm}

\begin{center}
{ \Large Weyl transverse diffeomorphism invariant theory of symmetric teleparallel gravity}
\vspace*{1.5cm}\\
{Yu Nakayama}
\end{center}
\vspace*{1.0cm}
\begin{center}

Department of Physics, Rikkyo University, Toshima, Tokyo 171-8501, Japan

\vspace{3.8cm}
\end{center}

\begin{abstract}
We construct a Weyl transverse diffeomorphism invariant theory of symmetric teleparallel gravity by employing the Weyl compensator formalism. The low-energy dynamics has a single spin two gravition without a scalar degree of freedom. By construction, it is equivalent to the unimodular gravity (as well as the Einstein gravity) at the non-linear level.
\end{abstract}

\thispagestyle{empty} 

\setcounter{page}{0}

\newpage

\section{Introduction}
While Einstein formulated a relativistic theory of gravity in terms of metric geometry, different formulations of gravitational theory may play prominent roles in theoretical physics. From the viewpoint of quantum gravity or string theory, a geometrical description may be just a low-energy emergent property of the fundamental quantum theory. Even at the technical level, going beyond the geometrical formulation is practically fruitful. For example, coupling to fermions requires the use of the tangent bundle through vielbeins. Torsion may appear naturally in supergravity in superspace. Furthermore, (super-)conformal approach to (super-)gravity is useful in black hole physics e.g. to study its entropy formula. 

The Einstein gravity is based on the connection that is torsion-free and metric compatible. The Weitzenb\"ock teleparallel gravity is based on the connection that is metric compatible and flat (but with torsion). See e.g. \cite{tele} for a review. The symmetric teleparallel gravity \cite{Nester:1998mp}\cite{BeltranJimenez:2017tkd}, which will be our main focus, is based on the connection that is torsion-free and flat (but not metric compatible). We know that these three formulations are equivalent \cite{Jimenez:2019woj}\cite{BeltranJimenez:2019odq}\cite{Bahamonde:2021gfp}\cite{Lu:2021wif} at least classically.

A theory of the metric geometry may admit further symmetry. For example, the Weyl symmetry demands that the theory must be invariant under the local rescaling of the metric: $g_{\mu\nu} \to e^{2\sigma} g_{\mu\nu}$. It means that the theory is scale-free. In the conventional approach in terms of the Riemann geometry, however, it is generally believed that the Weyl invariant action (i.g. Weyl tensor squared) in four-dimension predicts a propagating ghost-degree of freedom. It is possible to rescue the Weyl symmetry (without introducing a ghost) by abandoning the full diffeomorphism and considering a theory invariant only under the transverse diffeomprhism (i.e. the diffeomorphism whose determinant is unity $\det(\frac{\partial \xi^\mu}{\partial x^\nu}) = 1$). Such a theory is known as the unimodular gravity, which describes a sensible theory of gravity.

What will happen to the Weyl symmetry if we go beyond the Riemann geometry?
In this paper, we would like to construct a Weyl transverse diffeomorphism invariant theory of symmetric teleparallel gravity.
The Weyl transverse diffeomorphism is the only alternative symmetry (to the usual diffeomorphism) to realize a consistent theory of Fierz-Pauli theory of massless spin two particle. It consists of the Weyl transformation of the metric and the diffeomorphims whose determinant is unity.

Inspired by the linearized analysis, there is a natural ansatz for the action:
\begin{align}
S_W = \int d^4x {g}^{\frac{1}{4}} \left( c_1 Q_{\alpha \mu \nu} Q^{\alpha \mu \nu} + c_2 Q_{\alpha \mu \nu} Q^{\mu \alpha \nu} + c_3 Q_{\alpha} Q^{\alpha} + c_4 \bar{Q}_{\alpha} \bar{Q}^{\alpha} + c_5 \bar{Q}_\alpha Q^{\alpha} \right) \ , \label{ansatz}
\end{align}
where $Q_{\alpha \mu \nu} = \partial_\alpha g_{\mu\nu}$ is the non-metricity tensor in the coincident gauge (with $Q_\mu =  Q_{\mu \  \alpha}^{\ \alpha} $ and $\bar{Q}_{\mu} =  Q_{\ \alpha\mu }^{\alpha } $). The Weyl invariance demands $c_1 = \frac{1}{4}$, $c_2 + c_4= -\frac{1}{2}$, $c_3 = -\frac{3}{32}$ and $c_5= \frac{1}{4}$. See e.g. \cite{Alvarez:2006uu}\cite{Jimenez:2019woj} for a linearized theory. 

We, however, have some puzzles. The  Weyl transverse diffeomorphism invariant theory (as described by the above action) is not invariant under the full diffeomorphism (whose determinant is not unity). Then how can we demand the coincident gauge? We may also ask; how can we resolve the one parameter ambiguities of $c_2 + c_4 = -\frac{1}{2}$? We will answer these questions by employing the Weyl compensator formalism.

To anticipate our discussions below, we will argue that  ansatz \eqref{ansatz} is written in the coincident gauge, and the original diffeomorphism has been already fixed. We emphasize that the Weyl transverse diffeomorphism is the symmetry of the action in the coincident gauge. The transverse diffeomorphsim is distinguished from the original diffeomorphism that is fixed by taking the coincident gauge. This is similar to the symmetric telleparallel analogue of the Einstein gravity \cite{Jimenez:2019woj}, where we have the larger symmetry so that the diffeomorphism remains (or emerges) even in the coincident gauge.

In the pioneering work \cite{Gakis:2019rdd}, they constructed   a Weyl invariant gravity action in the symmetric teleparallel formalism. Their action contains $\int d^4x g^{\frac{1}{2}}Q_c^2$, where $Q_c = c_1 Q_{\alpha\mu\nu}Q^{\alpha\mu\nu} + \cdots$ and it has a higher derivative nature in the coincident gauge. Our strategy is different from theirs in the sense that we construct a different Weyl invariant action which becomes the unimodular gravity in the coincident gauge. It is only invariant under the volume-preserving diffeomorphism (i.e. the diffeomorphism whose Jacobian is unity), but it offers a consistent theory of gravity. The equivalence of the unimodular gravity and the Einstein gravity (with arbitrary cosmological constant) will be reviewed in the main text. 

Due to the Weyl invariance, our theory is scale-free and there is no dimensional parameter. Furthermore, the effective cosmological constant appears as an integration constant rather than a parameter in the theory. Many of these features are shared by the unimodular gravity, and our reformulation may be philosophically appealing although physical predictions are identical to the Einstein gravity as we will show.

\section{Weyl transverse diffeomorphism invariant theory of symmetric teleparallel gravity}
In the symmetric teleparallel formalism of gravity, we consider a general connection without torsion
\begin{align}
\Gamma^{\alpha}_{\ \mu\nu} = \Gamma^{\alpha}_{\ \nu\mu} \ ,
\end{align}
which is not necessarily metric compatible:
\begin{align}
Q_{\alpha \mu\nu} = D_\alpha g_{\mu\nu} \ ,
\end{align}
where the covariant derivative $D_\alpha$ is with respect to the torsion-free connection $\Gamma^{\alpha}_{\ \mu\nu} $.
The non-metricity tensor can be obtained from the connection as
\begin{align}
\Gamma^{\alpha}_{\ \mu\nu} = \left\{\right\}^{\alpha}_{\ \mu \nu} + L^{\alpha}_{\ \mu\nu} \ ,
\end{align}
where $\left\{\right\}^{\alpha}_{\ \mu \nu} = g^{\alpha \beta} \frac{1}{2}(\partial_\mu g_{\beta \nu} + \partial_\nu g_{\beta \mu} - \partial_\beta g_{\mu \nu}) $ is the Christoffel symbol  constructed out of the metric tensor $g_{\mu\nu}$, and $L^{\alpha}_{\  \mu\nu}$ is the so-called disformation tensor:
\begin{align}
L^{\alpha}_{\ \mu\nu} = \frac{1}{2}{Q}^{\alpha}_{\ \mu\nu} - Q_{(\mu \   \nu)}^{\ \ \alpha} \ .
\end{align}
To simplify the notation, we introduce $Q_\mu = Q_{\mu \  \alpha}^{\ \alpha} $ and $\bar{Q}_{\mu} =  Q_{\ \alpha\mu }^{\alpha }$ as the two independent contraction of the non-metricity tensor. As in general relativity, all the (tensor or non-tensor) indices are raised or lowered by the metric tensor $g_{\mu\nu}$, but we should be careful about the non-metricity nature of covariant derivatives.

Our theory of gravity is invariant under the Weyl transformation.\footnote{In the literature, it is also called ``conformal  transformation", but we preferably use the word Weyl transformation in this paper. It is also different from the scale transformation \cite{Nakayama:2013is}.} To understand the Weyl transformation of the connection, we note that  in the symmetric teleparallel gravity, we will impose the flatness condition of the connection. In the coincident gauge that we will discuss in a moment, we will impose the condition $Q_{\alpha \mu\nu} = \partial_\alpha g_{\mu\nu}$, and it is natural that both sides transform in the same manner. Motivated by this observation, we introduce the following (infinitesimal) Weyl transformation of various fields (see also \cite{Iosifidis:2018zwo}\cite{Gakis:2019rdd}):
\begin{align}
\delta_W g_{\mu\nu} &= 2\sigma g_{\mu\nu} \cr
\delta_W g & = 8\sigma g \cr
\delta_W \Gamma^{\alpha}_{\ \mu\nu} &= 0 \cr
\delta_W Q^{\alpha}_{\ \mu\nu} &= 2 g_{\mu\nu} g^{\alpha \beta} \partial_\beta \sigma \cr
\delta_W Q_\mu &= 8 \partial_\mu \sigma \cr
\delta_W \bar{Q}_\mu & = 2 \partial_\mu \sigma \cr
\delta_W \phi &= - \sigma \phi \ .  \label{Weyl}
\end{align}
Here $g = |\det g_{\mu\nu}|$ is the determinant of the metric; in this paper we work exclusively in four-dimensional space-time. For later purposes, we have included the Weyl transformation of the compensator scalar $\phi$. Here, we emphasize that the connection $\Gamma^{\alpha}_{\ \mu\nu}$ is invariant under the Weyl transformation (unlike the Christoffel symbol), which is similar to the Palatini formulation of the Weyl invariant theories of gravity \cite{Edery:2019txq}.

We start with the Weyl compensated form of the symmetric teleparallel gravity. The action is given by
\begin{align}
S = \int d^4x \sqrt{g} \left( \phi^2 (Q_E + \mathcal{D}_\alpha(Q^\alpha -\bar{Q}^\alpha)) - 6 g^{\mu\nu} \partial_\mu \phi \partial_\nu \phi + \lambda^{\mu\nu\rho\sigma} R_{\mu\nu\rho\sigma} \right)  \ . \label{start}
\end{align}
Here, $\mathcal{D}_\alpha$ is the metric compatible covariant derivative with respect to the Christoffel symbol but not with respect to $\Gamma^{\alpha}_{\ \mu\nu} $. If we were not happy with this choice of the connection in the covariant derivative, following a spirit of the teleparallel formulation, we could do an integration by part and use the ordinary derivative acting on $\phi^2$ and we could remove the reference to the metric compatible covariant derivative $\mathcal{D}_\alpha$ from the action.\footnote{Throughout this paper, we neglect surface terms.} The curvature tensor $R^{\rho}_{\ \sigma \mu \nu} = \partial_\mu \Gamma^{\rho}_{\ \nu \sigma} -\partial_\nu \Gamma^{\rho}_{\ \mu\sigma} + \Gamma^{\rho}_{\ \mu \lambda} \Gamma^{\lambda}_{\ \nu \sigma}- \Gamma^{\rho}_{\ \nu \lambda} \Gamma^{\lambda}_{\ \mu \sigma} $ here is constructed out of the connection $\Gamma^{\alpha}_{\ \mu\nu}$, and we have introduced the Lagrange multiplier $\lambda^{\mu\nu\rho\sigma}$ in order to impose the flatness condition on $\Gamma^{\alpha}_{\ \mu\nu}$.
 We will treat $g_{\mu\nu}$, $\Gamma^{\alpha}_{\ \mu\nu}$, $\phi$ and $\lambda^{\mu\nu\rho\sigma}$ as independent fields (see also \cite{Hohmann:2021fpr} for a variational perspective). 

The quadratic term ${Q}_E$ is given by
\begin{align}
{Q}_E = c_1 Q_{\alpha \mu \nu} Q^{\alpha \mu \nu} + c_2 Q_{\alpha \mu \nu} Q^{\mu \alpha \nu} + c_3 Q_{\alpha} Q^{\alpha} + c_4 \bar{Q}_{\alpha} \bar{Q}^{\alpha} + c_5 \bar{Q}_\alpha Q^{\alpha}
\end{align}
with a particular choice of the parameters: $c_1 = \frac{1}{4}$, $c_2 = -\frac{1}{2}$, $c_3 = -\frac{1}{4}$, $c_4 =0$ and $c_5 = \frac{1}{2}$. 
The action is invariant under the  diffeomorphism as well as the Weyl transformation introduced in \eqref{Weyl}. Note that even without specifying $c_i$ with the above choice, the action is invariant under diffeomorphism, assuming that the non-metricity is a tensor under the coordinate transformation, but our particular choice of $c_i$ will show an enhanced symmetry in the coincident gauge (in addition to the Weyl symmetry). 

Before we  proceed, let us briefly recall that this theory is equivalent to the Einstein gravity. We can use the Weyl symmetry to fix $\phi=1$. From the flatness condition $R_{\mu\nu\rho\sigma} = 0$, the connection $\Gamma^{\alpha}_{\ \mu\nu}$ is flat and hence locally a pure gauge. We can then impose the coincident gauge 
 $Q_{\alpha \mu\nu} = \partial_\alpha g_{\mu\nu}$ by setting $\Gamma^{\alpha}_{\ \mu\nu}=0$ from the diffeomorphism. When the curvature scalar $R$ constructed out of $\Gamma^{\alpha}_{\ \mu\nu}$ vanishes, we have the identity
  \begin{align}
-\mathcal{R} =  \frac{1}{4} Q_{\alpha \mu \nu} Q^{\alpha \mu \nu} -\frac{1}{{2}} Q_{\alpha \mu \nu} Q^{\mu \alpha \nu} -\frac{1}{4} Q_{\alpha} Q^{\alpha} + \frac{1}{2} \bar{Q}_\alpha Q^{\alpha} + \mathcal{D}^\mu(Q_\mu -\bar{Q}_\mu)  \ ,
\end{align}
where $\mathcal{R}$ is the Ricci scalar with respect to the Christoffel symbol. Therefore the original action with the Weyl gauge fixing $\phi=1$ is equivalent to the Einstein gravity:
 \begin{align}
S_{E} = -\int d^4x \sqrt{g} \mathcal{R} \label{Einstein}
 \end{align}
 up to surface terms. Alternatively, one may substitute the coincident gauge $Q_{\alpha \mu\nu} = \partial_\alpha g_{\mu\nu}$ condition into $Q_E$ and obtain the Einstein action up to surface terms.

Note that with our particular choice of parameters $c_i$, action \eqref{Einstein} is invariant under the emergent diffeomorphism even though we have already fixed the original diffeomorphism   in the coincident gauge (see e.g. \cite{Jimenez:2019woj} for more details on this point). An implicit assumption in the symmetric teleparallel formulation of gravity is to regard this emergent diffeomorphism as a gauge symmetry, meaning we physically identify solutions that are related by the emergent diffeomorphism. If we did not do so, we would end up with infinitely many redundant solutions for the original variables.

Now we would like to construct a Weyl transverse diffeomorphism invariant theory of symmetric teleparallel gravity from the same action but with a different gauge fixing. For this purpose, we again fix the original diffeomorphism to achieve the coincident gauge $Q_{\alpha \mu\nu} = \partial_\alpha g_{\mu\nu}$  from the flatness condition on $\Gamma^{\alpha}_{\ \mu\nu}$. As we have seen, the gauge fixed action has the Weyl symmetry as well as the emergent diffeomorphism  (which we stress is distinct from  the original diffeomorphism we have already fixed). Now, instead of fixing the Weyl symmetry, we fix the ``volume-changing" diffeomorphism (or conformal diffeomorphism) by setting $\phi = g^{-\frac{1}{8}}$. This condition is invariant under the Weyl transformation \eqref{Weyl}, so we have to use diffeomorphism to attain the gauge condition. 

After these gauge fixings, up to surface terms the resultant action is now given by 
\begin{align}
S_W = \int d^4x {g}^{\frac{1}{4}} {Q}_W \ , 
\end{align}
where 
\begin{align}
{Q}_W = c_1 Q_{\alpha \mu \nu} Q^{\alpha \mu \nu} + c_2 Q_{\alpha \mu \nu} Q^{\mu \alpha \nu} + c_3 Q_{\alpha} Q^{\alpha} + c_4 \bar{Q}_{\alpha} \bar{Q}^{\alpha} + c_5 \bar{Q}_\alpha Q^{\alpha} \label{final}
\end{align}
with $c_1 = \frac{1}{4}$, $c_2= -\frac{1}{2}$, $c_3 = -\frac{3}{32}$, $c_4=0 $ and $c_5= \frac{1}{4}$. Note that the coefficients $c_3$ and $c_5$ got extra contribution from $g^{\frac{1}{4}} \mathcal{D}_\alpha (Q^\alpha -\bar{Q}^\alpha)$ in \eqref{start} by performing the integration by part since it is no longer a total derivative.\footnote{This  parameter is a particular case of three parameter families studied in \cite{Gakis:2019rdd}. The importance of this particular value, naturally chosen in our approach, will be further discussed in the last paragraph of the section.} This action is invariant under the Weyl symmetry $g_{\mu\nu} \to e^{2\sigma} g_{\mu\nu}$ and the (emergent) transverse diffeomorphism (or volume preserving diffeomorphism, i.e. the diffeomorphism whose determinant is unity $\det(\frac{\partial \xi^\mu}{\partial x^\nu}) = 1$) in the coincident gauge. We have put the adjective ``emergent" here in order to emphasize it is the symmetry in the coincident gauge and it should be distinguished with the original diffeomorphism which we have already fixed by attaining the coincident gauge.

By construction, this action is equivalent to the unimodular gravity. To see this explicitly, we realize that up to surface terms the action can be rewritten as \begin{align}
    S_W = -\int d^4x g^{\frac{1}{4}} \left( \mathcal{R} + \frac{3}{32} g^{-2} g^{\mu\nu} \partial_\mu g \partial_\nu g \right) \ , 
\end{align}
which describes the unimodular gravity \cite{Alvarez:2006uu}\cite{Alvarez:2010cg}\cite{Oda:2016pok} (up on fixing the Weyl symmetry to set $g = 1$). Since we have started with the symmetric teleparallel formulation of the Einstein gravity, it is clear in our formulation that the Einstein gravity  is equivalent to the unimodular gravity (without matter at this point). See e.g. \cite{Alvarez:2005iy} for the review on the (classical) equivalence between the Einstein gravity and the unimodular gravity. 

We have a couple of comments about our theory.

Firstly, we could have started with the Einstein gravity with a cosmological constant by adding $\int d^4x \sqrt{g} \Lambda \phi^4$ to  the action \eqref{start} that we started with. If we fix the (emergent)  volume-changing diffeomorphism (i.e. a part of the diffeomorphism that can be cancelled by the Weyl transformation) by setting $\phi = g^{-\frac{1}{8}}$, the cosmological constant term becomes independent of any field variables and does not affect the equations of motion. This explains the fact that in the unimodular gravity, the cosmological constant is an ``integration constant" that cannot be determined from the equations of motion \cite{Henneaux:1989zc}. In other words, the unimodular gravity is equivalent to the Einstein gravity with an arbitrary cosmological constant; it is not one-to-one but one-to-many.

Secondly, we may linearize our final action \eqref{final} in the coincident gauge around the Minkowski space background $g_{\mu\nu} = \eta_{\mu\nu}$. The resulting linearized action is
\begin{align}
S_W = \int d^4x \left( \frac{1}{4} \partial_\alpha h_{\mu\nu} \partial^\alpha h^{\mu\nu} -\frac{1}{2} \partial_\mu h^{\mu\alpha} \partial^\nu h_{\nu\alpha} + \frac{1}{4}\partial_\mu h \partial_\nu h^{\nu\mu} -\frac{3}{32}\partial_\mu h \partial^\mu h  \right) \ , 
\end{align}
where $h = \eta^{\mu\nu} h_{\mu\nu}$.
As we know in the literature \cite{Alvarez:2006uu}, it describes a spin two massless graviton without a scalar degree of freedom. The linearized action can be expressed only in terms of  the traceless mode $\hat{h}_{\mu\nu} = h_{\mu\nu} - \frac{\eta_{\mu\nu}}{4} h $.  
It has the Weyl transverse diffeomorphism symmetry and it is known that this version of the Fierz-Pauli theory is ghost-free at the linearized level. 

Thirdly, we would like to discuss the degeneracy of $c_2+c_4$ \cite{BeltranJimenez:2019odq}. If our primary focus were the Weyl symmetry, there would exist one parameter ambiguities in $Q_W$. As mentioned in the introduction, as long as $c_2 + c_4 = -\frac{1}{2}$  (with all the other parameters are fixed as in \eqref{final}), it is Weyl invariant. In the linearized theory, only the combination of $c_2 + c_4$ appears because they are related by partial integration in the Minkowski space-time at the linearized level (but it is not so at the full non-linear level). Our theory, however, resolves the degeneracy by specifying $c_4 = 0$. We believe this is necessary in order to preserve the (emergent) diffeomorphism at the non-linear level to avoid any pathology. We currently do not know if $c_4\neq 0$ is actually pathological, but it is probable that at the non-linear level, the ghost-like degrees of freedom appear \cite{Boulware:1972yco}.

\section{Discussions}
In this paper, we have constructed a Weyl transverse diffeomorphism invariant theory of symmetric teleparallel gravity by employing the Weyl compensator formalism. By construction, it is equivalent to the unimodular gravity as well as the Einstein gravity at the non-linear level. 

We have demonstrated the equivalence without introducing matter. Now we want to discuss the effect of the matter. We first realize that it is always possible to make most of the bosonic matter action Weyl invariant by using the compensator $\phi$. Familiar gauge theories such as the Maxwell theory are already Weyl invariant and we do not need a compensator.  For a single scalar field $\Phi$ with a canonical kinetic term, we may consider the Weyl invariant action
\begin{align}
S_m = \int d^4x \sqrt{g} \left(- \phi^{2} g^{\mu\nu} \partial_\mu \Phi \partial_\nu \Phi - \phi^{4} V(\Phi) \right)  \ ,
\end{align}
where we assume that a scalar field $\Phi$ is invariant under the Weyl transformation. 

In the coincident gauge, the matter action has the emergent diffeomorphism.\footnote{If the matter action requires covariant derivatives, we have to carefully discuss which covariant derivative (i.e. metric compatible one or metric non-compatible one) is to be used. Depending on the choice, the emergent diffeomorphism could be lost and the equivalence to the Einstein theory becomes non-trivial.}
In order to use this action in our Weyl transverse diffeomorphism invariant theory of symmetric teleparallel gravity, we can fix the emergent volume-changing diffeomorphism by setting $\phi = g^{-\frac{1}{8}}$ as before.

From this perspective, the Einstein gravity coupled with any (bosonic) matter can be reformulated in the unimodualr gravity in our symmetric teleparallel formulation. We should, however, note that deriving the equations of motion in the unimodular gauge $g = 1$ requires a care. The matter coupling is not from the conventional energy-momentum tensor. The metric variation should be carefully taken with the condition $\phi = g^{-\frac{1}{8}}$. Otherwise, the equations of motion becomes inconsistent.

Introducing fermionic matter forces us to use the tangent bundle and vielbeins as fundamental dynamical degrees of freedom.
The idea of the symmetric teleparallel gravity should be reconsidered to take into account the vielbeins and spin connections. In the usual vielbein formalism, the metric is constructed out of the vielbein $e^a_\mu$ as $g_{\mu\nu} = e^a_\mu e^b_\nu \eta_{ab}$ and we assume that theory is invariant under the local Lorentz transformation acting on the tangent bundle index $a$. This formulation makes it possible to define a spinor field. For this purpose, we need to introduce the spin connection $\omega^{\mu}_{ab}$. In the usual formulation, we assume that the spin connection is compatible with the vielbein $D_\mu e^a_\nu = Q^a_{\mu\nu} = 0 $, but in the symmetric teleparallel formalism, we expect that $Q^{a}_{\mu\nu}$ becomes a dynamical degree of freedom. 
This is beyond the scope of this work but may worth studying further. 

Another interesting direction is to study the Weyl transverse invariant theory of the $f(Q)$  symmetric teleparallel gravity  \cite{Nester:1998mp}\cite{Anagnostopoulos:2021ydo}. The idea is to start with the general $f(Q)$ symmetric teleparallel gravity and make it Weyl invariant by using the compensating scalar $\phi$. Then in the gauge $\phi = g^{-\frac{1}{8}}$, one may obtain the Weyl transverse diffeomprhism invariant theory corresponding to the $f(Q)$ symmetric teleparallel gravity.\footnote{In the special case of $f(Q)$ theory studied in \cite{Gakis:2019rdd}, we do not need the compensator, so our construction gives the identical theory as theirs.} The physical consequence of such theories must be studied further in detail.

Some of the recent works on the applications of symmetric teleparallel gravity can be found in \cite{Harko:2018gxr}\cite{Hohmann:2018wxu}\cite{Soudi:2018dhv}\cite{Lu:2019hra}\cite{BeltranJimenez:2019tme}\cite{Lazkoz:2019sjl}\cite{Xu:2019sbp} \cite{Barros:2020bgg}\cite{Xu:2020yeg}\cite{Pradhan:2020brv}\cite{Mandal:2020buf}\cite{Zia:2021vhr}\cite{Khyllep:2021pcu}\cite{Frusciante:2021sio}\cite{Zhao:2021zab}\cite{Najera:2021puo}\cite{Lin:2021uqa}\cite{Pradhan:2021qnz}\cite{Bahamonde:2021gfp}\cite{Pati:2021ach}\cite{Beh:2021wva}\cite{Rudra:2021ksp}\cite{Esposito:2021ect}\cite{Agrawal:2021rur}\cite{Li:2021mdp}.

\section*{Acknowledgements}
I would like to thank W. D. Linch for informing me of the teleparallel gravity when I visited University of Maryland in 2013. After eight years, I still do not know if the teleparallel gravity can give a non-linear completion of the Virial supergravity \cite{Buchbinder:2002gh}\cite{Gates:2003cz}\cite{Nakayama:2014kua}. This work is in part supported by JSPS KAKENHI Grant Number 21K03581.

\end{document}